\documentclass{elsart}

\usepackage{amssymb}

\begin{document}
\begin{frontmatter}

\title{When non-extensive entropy becomes extensive}

\author[wada]{Wada Tatsuaki\corauthref{cor1}} and
\ead{wada@ee.ibaraki.ac.jp}
\author[saito]{Saito Takeshi\thanksref{add}}
\ead{saito@kif.co.jp}
\address[wada]{Department of Electrical and Electronic Engineering, 
Ibaraki University, Hitachi,~Ibaraki, 316-8511, Japan}
\corauth[cor1]{Corresponding author.}
\address[saito]{Complex Functional Robot Laboratories, 
Graduate School of Science and Engineering, 
Ibaraki University, Hitachi, Ibaraki, 316-8511, Japan}
\thanks[add]{Present address: KIF \& Co., Ltd., Tokyo Dia Bldg. \#5, 
15 floor, 1-28-23 Shinkawa Chuo-ku, Tokyo, 104-0033, Japan}

\begin{abstract}
Tsallis' non-extensive entropy $S_q$ enables us to treat both a power and
exponential evolutions of underlying microscopic dynamics 
on equal footing by adjusting the variable entropic index $q$ to
proper one $q^*$.
We propose an alternative constraint of obtaining the proper entropic 
index $q^*$ that the non-additive
conditional entropy becomes additive if and only if $q=q^*$
in spite of that the associated
system cannot be decomposed into statistically independent subsystems.
Long-range (time) correlation expressed by $q$-exponential function
is discussed based on the nature that $q$-exponential function cannot
be factorized into independent factors when $q \ne 1$.
\end{abstract}

\end{frontmatter}

\section{Introduction}
There has been growing interest in the non-extensive statistical
mechanics 
\cite{Tsal88,Tsal99} based on Tsallis' generalized entropy 
(in $k_{\rm B}=1$ unit):
\begin{equation}
  S_q = \frac{1 - \sum_i p_i^q}{q - 1}.
\hspace{1cm}(\sum_i p_i=1; \hspace{3mm} q\in {\mathcal R})
\label{Sq}
\end{equation}

At least formally, Tsallis' entropy is an extension of 
conventional Boltzmann-Shannon (BS) entropy with one-real-parameter of $q$.
In the limit of $q \to 1$, Tsallis' entropy Eq. (\ref{Sq}) reduces to
BS entropy, $S_1 \equiv  -\sum_i p_i \ln p_i$,
since $p_i^{q-1} = e^{(q-1) \ln p_i} \approx 1+(q-1) \ln p_i$.

The parameter $q$ may be interpreted as a quantity characterizing 
the degree of non-extensivity of Tsallis' entropy through 
the so-called {\it pseudo-additivity}:
\begin{equation}
  S_q(A, B) =  \quad S_q(A) + S_q(B) + (1-q) S_q(A) S_q(B),
  \label{rel:p_add}
\end{equation}
\noindent where $A$ and $B$ denote two statistically 
independent sub-systems.
It is worth while to realize that the pseudo-additivity of $S_{q}$ is one of 
the crucial ingredients of Tsallis' non-extensive statistical mechanics. 
In fact, the uniqueness of Tsallis' entropic form is proved \cite{unique} 
for an entropy that fulfills the generalization of 
the Shannon-Kinchin axioms \cite{SK-theorem} based on the pseudo-additive
conditional entropy obeying the pseudo-additivity instead of additivity.
Then what is a role of the pseudo-additivity?
By rewriting Eq. (\ref{Sq}) as the following form,
\begin{equation}
  S_q = -\sum_i p_i^q \ln_q p_i,
\end{equation}
\noindent
we see that the pseudo-additivity of Tsallis' entropy comes from
the $q$-logarithmic function, which is defined by
\begin{equation}
  \ln_q(x) \equiv \frac{x^{1-q} - 1}{1-q},
\end{equation}
\noindent
since it equips the pseudo-additivity as
\begin{equation}
 \ln_q(x y) = \ln_q x + \ln_q y + (1-q) \ln_q x \ln_q y.
\end{equation}

The inverse function of the $q$-logarithmic function is $q$-exponential one, 
which is defined by
\begin{equation}
  \exp_q(x) \equiv [1+(1-q)x]^{\frac{1}{1-q}},
\end{equation}
\noindent
for $1+(1-q)x > 0$ and otherwise $\exp_q(x) = 0$.

As Tsallis \cite{Tsal00} has already pointed out, the parameter $q$ plays 
a similar role as the light velocity $c$ 
in special relativity or Planck's constant $\hbar$ in quantum 
mechanics  in the sense of a one-parameter extension of classical mechanics.  
Unlike  $c$ or $\hbar$, however, $q$ does not seem to be a universal constant.
Thus it is a natural question whether $q$ is merely an 
adjustable parameter or not in the non-extensive statistical mechanics.
In some cases the parameter $q$ has no physical meaning, but when
it is used as an adjustable parameter the resulting distributions give 
excellent agreement with experimental data. 
In other but a few cases \cite{Cost97,Lato99,Mour00,Buia99,Arim00}, 
$q$ is uniquely determined
by the constraints of the problem and thereby $q$ may have a physical meaning. 
Recent studies of the characterization of mixing
in one-dimensional (1D) dissipative maps 
\cite{Cost97,Lato99,Mour00} and of symbolic sequences \cite{Buia99} seems to 
provide a positive answer to the above question since there exists
the special value $q^*$ such that $S_{q^*}$ becomes linear. 
For example,
in the studies \cite{Cost97,Lato99} of mixing in simple logistic map, 
$q^*$ can be obtained by three different methods based on: i) the upper bound 
of a power-law sensitivity $\xi$ to initial conditions; ii) the singularity 
indices in multi-fractal structure; and iii) the rate of information loss 
in terms of $S_q$. The remarkable fact is that all methods lead to the same 
value of $q^* \simeq 0.24$, which may shed some light on the physical 
meaning of $q$.
They established some 
connections among the sensitivity $\xi$ to initial conditions, Tsallis'
 entropy $S_q$ and the proper 
entropic index $q^*$.  In particular we focus on the work of Buiatti 
{\it et~al.} \cite{Buia99}, in which they have shown, for the symbolic 
sequences with length of $N$, that the generalized block entropy 
\cite{block-entropy} is 
proportional to 
$N$ when the proper entropic index $q^{*}$ is used.
In other words, there may exist a proper entropic index $q^{*}$, 
which may statistically characterize a non-extensive system.

In this work we study a reason why the generalized block entropy in the work
of Buiatti {\it et~al.} \cite{Buia99} 
is proportional to the length $N$ of symbolic sequences  when 
we use the proper entropic index $q^{*}$.  
In particular we focus our attention on a 
role of the pseudo-additivity of the conditional entropy in characterizing 
a non-extensive system with the proper entropic index $q^*$. 
We reformulate the 
constraint of obtaining the proper entropic index $q^{*}$ as follows:
the pseudo-additive conditional entropy becomes additive with 
respect to $N$ when the proper entropic index $q^{*}$ is used. In other words,
for the special value $q^*$ of entropic index, the additivity of 
the conditional 
entropy is held in spite of that the involved subsystems are not
statistically independent of each other.

The rest of the paper is organized as follows: 
in the next section we explain
the constraint of obtaining the proper entropic index $q^*$ in 
the work \cite{Buia99} of Buiatti {\it et~al.}, and propose our constraint 
of obtaining $q^*$.
We then show the equivalence of the two constraints and discuss the 
underlying simple mechanism of why the 
conditional entropy becomes additive for the proper entropic index $q^*$ 
under the assumption of equi-probability. 
In Section 3, we discuss long-range correlation expressed by
$q$-exponential function. Section 4 is devoted to our conclusions.

\section{How to determine a proper entropic index}
Buiatti {\it et~al.} \cite{Buia99} showed, by studying a symbolic binary 
sequence $\{ \sigma_1, \sigma_2, \cdots \}$ with a long-range 
correlation, that for the probability $p(\sigma_1, 
\cdots, \sigma_N)$ of each path with length of $N$, the generalized 
block entropy \cite{block-entropy},
\begin{equation}
 S_q(N) \equiv \frac{1-\sum_{\sigma_1, \cdots, \sigma_N}
 p(\sigma_1, \cdots, \sigma_N)^{q}}{q-1},
\end{equation}
\noindent is 
proportional to $N$ if and only if the variable index $q$ equals 
the proper entropic index $q^*$.

We reformulate this constraint as the following.
The pseudo-additive conditional entropy \cite{Abe00}, which is defined by
\begin{equation}
  S_q(N \vert 1) 
  \equiv \frac{S_q(N+1) - S_q(1)}{1 + (1-q) S_q(1)},
  \label{cond-ent}
\end{equation}
should satisfy the {\it additivity},
\begin{equation}
  S_q(N \vert 1) = S_q(N-1 \vert 1) + S_q(1 \vert 1),
  \label{additive-Sq}
\end{equation}
when $q$ is equal to the proper entropic index $q^{*}$.
At first sight our constraint seems to be paradoxical,
since $S_q$ is pseudo-additive in general. We explain in 
the followings that the equivalence of the two constraints which determine
$q^*$, and discuss an underlying simple mechanism connecting the variable 
entropic index $q$ with the proper index $q^*$. 

\subsection{Proof of the equivalence of the two constraints}
The method of Buiatti {\it et~al.} is rephrased as follows: $S_q(N)$ is 
a linear function of $N$ when $q=q^*$, i.e.  
\begin{equation}
   S_{q^*}(N) = (N-1) L_{q^*} + S_{q^*}(1),
\label{lin-Sq} 
\end{equation}
\noindent
where $L_{q^*}$ is a proportional constant, or equivalently
\begin{equation}
   S_{q^*}(N+1) - S_{q^*}(N) = S_{q^*}(2) - S_{q^*}(1) = L_{q^*}, 
\hspace{8mm} {\rm for} 
   \hspace{2mm} N > 1.
   \label{linear-Sq}
\end{equation}

Subtracting $S_{q^*}(1)$ from the both sides and after a little bit algebra,
Eq. (\ref{linear-Sq}) is rewritten as
\begin{equation}
  S_{q^*}(N+1) - S_{q^*}(1) = S_{q^*}(N)- S_{q^*}(1) + S_{q^*}(2) - S_{q^*}(1).
  \label{linear-Sq2}
\end{equation}
\noindent
Dividing the both sides of Eq. (\ref{linear-Sq2}) by $1+(1-q^*)S_{q^*}(1)$ 
and using the definition of the 
conditional entropy of Eq. (\ref{cond-ent}), it is obvious that 
Eq. (\ref{linear-Sq}) is equivalent to Eq. (\ref{additive-Sq}).
Hence we have reformulated the method of Buiatti {\it et~al.} as 
follows: the conditional entropy becomes additive when
we use the proper entropic index $q^*$.      

\subsection{A reason why $S_{q^*}(N|1)$ is proportional to $N$}
Having explained the equivalence of Buiatti {\it et al.} and our methods of 
obtaining a proper entropic index $q^*$,
we now consider  why the conditional entropy $S_q(N|1)$ is proportional
to the length $N$ of symbolic sequences when $q=q^*$.

Under the assumption of equi-probability, 
Tsallis' entropy can be written in terms of the number of states $W(N)$ 
for the symbolic sequences with the length of $N$ as
\begin{equation}
  S_q(N) = \ln_q W(N).
  \label{equi-prob}
\end{equation}
Then the conditional entropy of Eq. (\ref{cond-ent}) is 
expressed in terms of $W(N)$ as
\begin{equation}
  S_q(N|1) = \ln_q\frac{W(N+1)}{W(1)}.
  \label{cond-Sq}
\end{equation}

Suppose that the number of states $W(N)$ obeys a power-law evolution,
which can be well described by
\begin{equation}
  W(N+1) = W(1) \exp_{q^*}( L'_{q^*} \; N ),
  \label{W}
\end{equation}
\noindent
with the proper $q^*$ of a system of interest, 
where $L'_{q^*}$ is another constant. 
The relation between $L'_{q^*}$ and $L_{q^*}$ of Eq.~(\ref{linear-Sq}) 
is discussed in the next section.
Substituting Eq.~(\ref{W}) into Eq.~(\ref{cond-Sq}), the corresponding 
conditional entropy can be written as
\begin{equation}
  S_q(N|1) = \ln_q [\frac{W(N+1)}{W(1)}] 
           = \ln_q[ \exp_{q^*}(L'_{q^*} \; N) ].
\end{equation}
\noindent
Now we readily see that 
  $S_q(N|1)$ is proportional to $N$, if and only if we set $q$ to $q^*$.
In other words, if $W$ obeys the $q^*$-exponential evolution of 
Eq. (\ref{W}), 
then it is reasonable
 to use its inverse function in order to define the conditional entropy.

\section{Long-range correlation expressed by $q$-exponential function}
Tsallis' entropic description may be well suited for a long-range
correlated system which obeys a power-law evolution described by 
$q$-exponential function.
Then how can $q$-exponential function express long-range correlation? 
We here explain that long-range correlation may be expressed by the 
non-factorizability of $q$-exponential function into independent terms.
The long-range correlation in this case means the initial condition 
dependency of long duration.
It is known that $q$-exponential function cannot be resolved into a product 
of independent terms unless $q=1$. For example $\exp_{q}(t + s)$ is not 
resolved into the independent factors as $\exp_{q}(x) \cdot \exp_{q}(s)$.
Instead it can be expressed as the product of the dependent factors as 
$\exp_{q}(x) \cdot \exp_{q}(s / \{1+(1-q)x \})$.

Let us focus on the long-range correlation associated with the power-law
evolution of $W(N)$ described by Eq. (\ref{W}).
Using Eqs. (\ref{lin-Sq}) and (\ref{equi-prob}), the $q$-exponential 
dependency of $W(N+1)$ can be expressed as
\begin{eqnarray}
 W(N+1)&=& \exp_q[S_q(N+1)] = \exp_q[ S_q(1) + L_q N] \nonumber \\
       &=& \exp_q[S_q(1)] \cdot \exp_q[\frac{L_q N}{1+(1-q)S_q(1)}] 
       \nonumber \\
       &=& \exp_q[S_q(1)] \cdot \exp_q[\frac{L_q}{1+(1-q)S_q(1)}]
       \nonumber \\
    &\times&  \cdots \times
       \exp_q [ \frac{L_q}{1 + (1-q)\{S_q(1)+N-1\}}].
    \label{long-range}
\end{eqnarray}
Note that $S_q(1)$ appears in all terms and this reflects the initial 
condition dependency of long duration.
This feature is consistent with the single-trajectory approach by Montangero
{\it et~al.} \cite{Mont99} in which they fix a given initial condition in
order to obtain the $q^*$ of the non-extensive version of Kolmogorov-Sinai
entropy for the dynamics of the logistic map at the chaotic threshold.
Because of the initial condition dependency of long duration, an averaging 
over many different initial conditions is not appropriate.

Now let us focus on the relation between the proportional constants
$L'_{q^*}$ and $L_{q^*}$ in the previous section. From the second line
of Eq. (\ref{long-range}), we see that
\begin{equation}
 W(N+1) = W(1) \cdot \exp_q[\frac{L_q N}{1+(1-q)S_q(1)}].
\end{equation}
\noindent
Comparing this with Eq. (\ref{W}), $L'_{q^*}$ and $L_{q^*}$ are related by
\begin{equation}
  L'_{q^*} = \frac{L_{q^*}}{1+(1-q)S_{q^*}(1)} = \frac{L_{q^*}}{W(1)^{1-q}}.
\end{equation}
\noindent
which is the same relation \cite{OLM} of $\lambda'= \lambda / \bar{Z}_q^{1-q}$
  between the Lagrange multiplier $\lambda'$ 
of optimal Lagrange multipliers (OLM) method and that $\lambda$ of 
Tsallis-Mendes-Plastino one in canonical ensemble formalism, where $\bar{Z}_q$
denotes partition function.

\section{Conclusions}
  We have proposed a constraint of obtaining the proper Tsallis' entropic
index $q^*$ in describing the evolutions of correlated symbolic 
sequences with length $N$. The proper entropic index $q^{*}$ can be 
determined by requiring that the conditional entropy $S_q(N|1)$ should
be proportional to $N$ if and only if $q$ equals the proper entropic index 
$q^*$. In other words $S_q$ becomes {\it additive} for the proper $q^*$.  
It is the non-factorizability of $q$-exponential function 
into independent terms that can express a long-range correlation.

\section*{Acknowledgments}
One of the authors (T.~W) acknowledges S.~Abe, T.~Arimitsu and N.~Arimitsu 
for useful comments and valuable discussion at the 9th symposium 
on Non-Equilibrium Statistical Physics held at Tsukuba, Japan.

\end{document}